\documentclass{article}
\usepackage{spconf,amsmath,graphicx,hyperref}
\usepackage{textcase}
\usepackage{enumitem}
\usepackage[font=small,labelfont=bf]{caption}
\usepackage{array}
\newcolumntype{C}[1]{>{\centering\arraybackslash}p{#1}}
\usepackage{subcaption} 

\makeatletter
\renewcommand\normalsize{%
	\@setfontsize\normalsize{9.7pt}{11.8pt}%
	\abovedisplayskip 8pt plus2pt minus5pt%
	\abovedisplayshortskip 0pt plus3pt%
	\belowdisplayshortskip 4pt plus3pt minus3pt%
	\belowdisplayskip \abovedisplayskip%
}
\normalsize

\renewenvironment{thebibliography}[1]
{\begin{oldthebibliography}{#1}%
		\setlength{\parskip}{0pt}      
		\setlength{\itemsep}{0.3ex}    
		\setlength{\parsep}{0pt}}      
	{\end{oldthebibliography}}
\makeatother

\title{Multimodal Deep Learning Method for Real-Time Spatial Room Impulse Response Computing}

\name{Zhiyu Li, Xinwen Yue, Shenghui Zhao, Jing Wang$^*$\thanks{$^*$Corresponding author's email: wangjing@bit.edu.cn}}
\address{School of Information and Electronics, Beijing Institute of Technology, Beijing, China}
\begin{document}
	\maketitle
	\renewcommand{\thefootnote}{} 
	\footnotetext{This work was supported by XXX}
	\renewcommand{\thefootnote}{\arabic{footnote}} 

\begin{abstract}\label{sec:abstract}
We propose a multimodal deep learning model for VR auralization that generates spatial room impulse responses (SRIRs) in real time to reconstruct scene-specific auditory perception. Employing SRIRs as the output reduces computational complexity and facilitates integration with personalized head-related transfer functions. The model takes two modalities as input: scene information and waveforms, where the waveform corresponds to the low-order reflections (LoR). LoR can be efficiently computed using geometrical acoustics (GA) but remains difficult for deep learning models to predict accurately. Scene geometry, acoustic properties, source coordinates, and listener coordinates are first used to compute LoR in real time via GA, and both LoR and these features are subsequently provided as inputs to the model. A new dataset was constructed, consisting of multiple scenes and their corresponding SRIRs. The dataset exhibits greater diversity. Experimental results demonstrate the superior performance of the proposed model.
\end{abstract}

\vspace{-1mm}
\section{Introduction}
\label{sec:intro}\vspace{-1mm}

Auralization in virtual reality (VR) is crucial for enhancing the sense of presence \cite{larsson2002better}. It refers to modeling the sound field of a scene so that the sound of a source becomes perceptible. Since VR scenes are inherently interactive, auralization must respond in real time to user actions. A common approach is to compute the room impulse response (RIR) in real time and convolve it with the source signal. The RIR acts as the system function between source and listener, from which the human auditory system can infer spatial and environmental information \cite{schutte2019percept}. Because of the binaural effect, binaural RIRs (BRIRs) are commonly used.

Geometrical acoustics (GA) is a mainstream method for real-time RIR (including BRIR) computation \cite{thery2019auralization}, which approximates sound propagation by modeling sound waves as rays emitted from the source. Since sound waves are minimally absorbed, accurate simulation often requires more than 20th-order reflections, and scattering further causes ray splitting, making the computational complexity of GA still challenging. In practice, the reflection order is usually reduced, sacrificing fidelity for efficiency. Deep learning (DL) methods have been proposed to address this challenge, offering promising avenues for improving VR auralization. Nevertheless, existing approaches face several limitations: \textbf{1)} RIR type: inherent constraints of monaural RIRs (MRIRs) and BRIRs; \textbf{2)} Datasets: current datasets are not fully aligned with VR auralization requirements; \textbf{3)} Model performance: further improvements remain necessary.

\textbf{Our contributions are as follows:}
\begin{itemize}[noitemsep, topsep=0pt, leftmargin=2em]
	\item Introduces a new task: real-time computation of spatial RIRs (SRIRs) using a DL-based model. Compared to MRIRs and BRIRs, SRIRs offer clear advantages, as detailed in Sec.~\ref{sec:Re_RIR}.
	
	\item Constructs a new dataset for VR auralization, containing over 1000 3D scenes, each with 1000 SRIRs corresponding to different source and listener positions.
	
	\item Proposes a new deep learning model that incorporates low-order reflections (LoR) as auxiliary modality, enabling real-time SRIR computation with superior performance.
\end{itemize}\vspace{-1mm}

\vspace{-1mm}
\section{Related Works}
\label{sec:Related}\vspace{-1mm}

\subsection{MRIR, BRIR and SRIR}
\label{sec:Re_RIR}\vspace{-1mm}

RIRs are typically divided into direct sound, early reflections, and late reverberation \cite{valimaki2012fifty}, each influencing auditory perception differently. The waveforms of the direct sound and early reflections provide source localization and width cues through binaural effect \cite{blauert1997spatial}\cite{lokki2011lateral}; the direct-to-reverberant energy ratio (DRR) conveys source distance \cite{bronkhorst1999auditory}; and late reverberation, via its coarse time-frequency characteristics \cite{alary2021perceptual}, contributes to envelopment and the perception of spatial extent. Key observations include: \textbf{1)} binaural effect is crucial for RIR perception; \textbf{2)} the waveforms of the direct sound and early reflections require accurate reconstruction, while late reverberation does not.

BRIRs and SRIRs preserve binaural effect, making them more suitable for VR auralization. SRIRs are multichannel RIRs that record RIRs from different directions on the surrounding sphere, and are often combined with Ambisonics \cite{merimaa2005spatial}. BRIRs vary with both listener position and orientation, whereas SRIRs vary only with position coordinates. SRIRs can be transformed into BRIRs using head-related transfer functions (HRTFs) \cite{hold2022parametric}, which eliminates the need for additional computation to account for head rotations, thereby reducing complexity while also providing a natural interface for incorporating personalized HRTFs. Personalized HRTFs can further enhance perceptual quality \cite{xie2020spatial}. Therefore, SRIRs offer substantial advantages as model outputs.

\begin{figure}[tbp]
	\centerline{\includegraphics[scale=0.5]{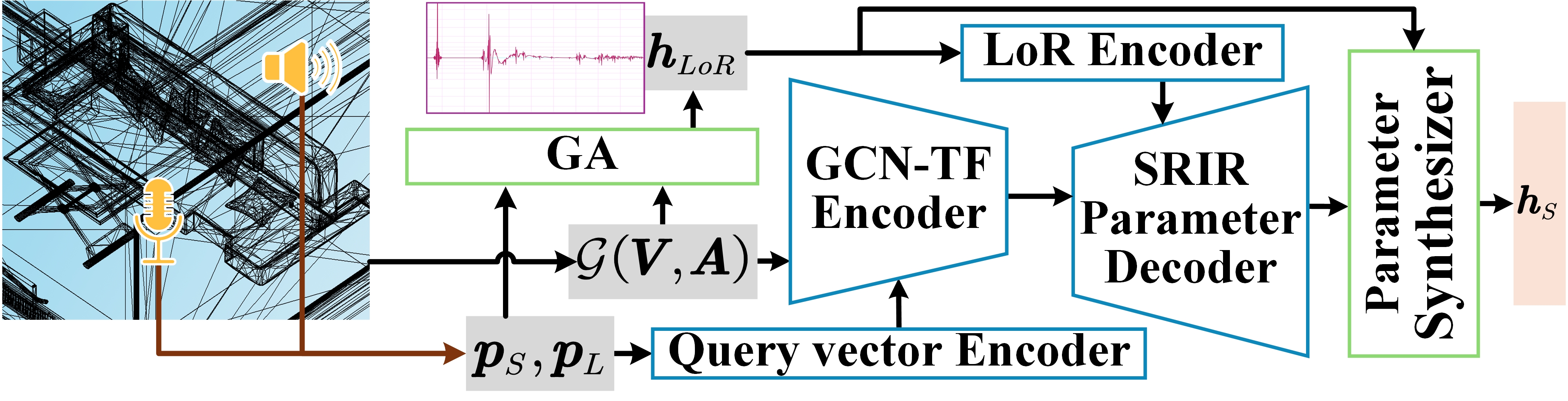}}
	\captionsetup{font=small, skip=2pt}
	\caption{Overall framework of the Multimodal DL-based model.}
	\label{model-total}
	\vspace{-3mm}
\end{figure}

\vspace{-1mm}
\subsection{Deep Learning Method}
\label{sec:Re_DL_model}\vspace{-1mm}

Several DL-based methods for RIR computation have been proposed. Neural Acoustic Fields (NAF)\cite{luo2022learning} achieve high performance but fail to generalize to unseen environments. Few-ShotRIR\cite{majumder2022few} estimates BRIRs for dynamic source-listener positions in unseen scenes using multiple environmental images and corresponding BRIRs. MESH2IR (M2R)\cite{ratnarajah2022mesh2ir} instead relies on 3D geometric scene models to generate MRIRs. Listen2Scene (L2S) \cite{ratnarajah2024listen2scene} extends M2R by incorporating scene acoustic properties, thereby generating better BRIRs. Building on M2R, M2PAIR \cite{li2025m2pair}\cite{li2024room} first predicts perceptual parameters of MRIRs and then synthesizes high-quality MRIRs. The xRIR\cite{liu2025hearing} extracts 3D information from depth images and generates MRIRs, while further enhancing RIR quality by employing a small set of known RIRs, predicting their weights to compute the target MRIR via weighted summation. 

Despite these advances, limitations remain. None of the above approaches support SRIR output, and while multimodal inputs often improve performance, models that take scene information as an input modality have not yet incorporated auxiliary modalities.
\vspace{-1mm}

\section{Our Approach}
\label{sec:method}\vspace{-1mm}

\subsection{Problem Formulation}
\label{sec:Problem}\vspace{-1mm}
We propose a scene-waveform multimodal deep learning approach for SRIR computation and design a model denoted as $\mathbf{F}$. The model takes as input the scene information (scene geometry, acoustic properties), source and listener coordinates, and the LoR waveforms corresponding to these coordinates. The scene geometry and acoustic properties are represented as a graph $\mathcal{G} \left( \boldsymbol{V},\boldsymbol{A} \right)$, where the vertex matrix $\boldsymbol{V}$ encodes all triangular faces, with each vertex vector $\boldsymbol{v}$ specifying position, shape, size, reflectivity, and scattering. The adjacency matrix $\boldsymbol{A}$ captures the connectivity among faces. Source and listener positions are represented as Cartesian coordinates $\boldsymbol{p}_S$ and $\boldsymbol{p}_L$, and the LoR is denoted as $\boldsymbol{h}_{LoR}$. The inference process for obtaining SRIR waveforms $\boldsymbol{h}_{S}$ is illustrated in Eq.~\eqref{F_func},
\vspace{-2mm}
\begin{equation}\boldsymbol{h}_S=\mathbf{F}\left( \mathcal{G} \left( \boldsymbol{V},\boldsymbol{A} \right) ,\boldsymbol{h}_{LoR},\boldsymbol{p}_S,\boldsymbol{p}_L \right) \label{F_func}
\vspace{-2mm}\end{equation}
where $\boldsymbol{h}_{LoR}$ is obtained by Eq.~\eqref{GA}.
\vspace{-2mm}
\begin{equation}\boldsymbol{h}_{LoR}=\mathrm{GA}\left( \mathcal{G} \left( \boldsymbol{V},\boldsymbol{A} \right) ,\boldsymbol{p}_S,\boldsymbol{p}_L,n_O \right) \label{GA} \vspace{-1mm}
\end{equation}

LoR denotes the first $n_O$-order reflections of the SRIR (here, $n_O=2$), in contrast to early reflections defined by a temporal boundary. LoR is chosen as a model input because the computational cost of GA increases exponentially with reflection order, whereas LoR can be computed efficiently. Furthermore, human auditory perception is particularly sensitive to LoR \cite{blauert1997spatial}, making it challenging for deep learning models to achieve high accuracy in this range. Therefore, we directly compute LoR using GA and incorporate it as an auxiliary modality.

\vspace{-1mm}
\subsection{Model Architecture}
\label{sec:model}\vspace{-2mm}
The overall architecture of the proposed model is illustrated in Fig.~\ref{model-total}. The primary modules and their detailed structures are described as follows:

\begin{figure}[tbp]
	\centerline{\includegraphics[scale=0.58]{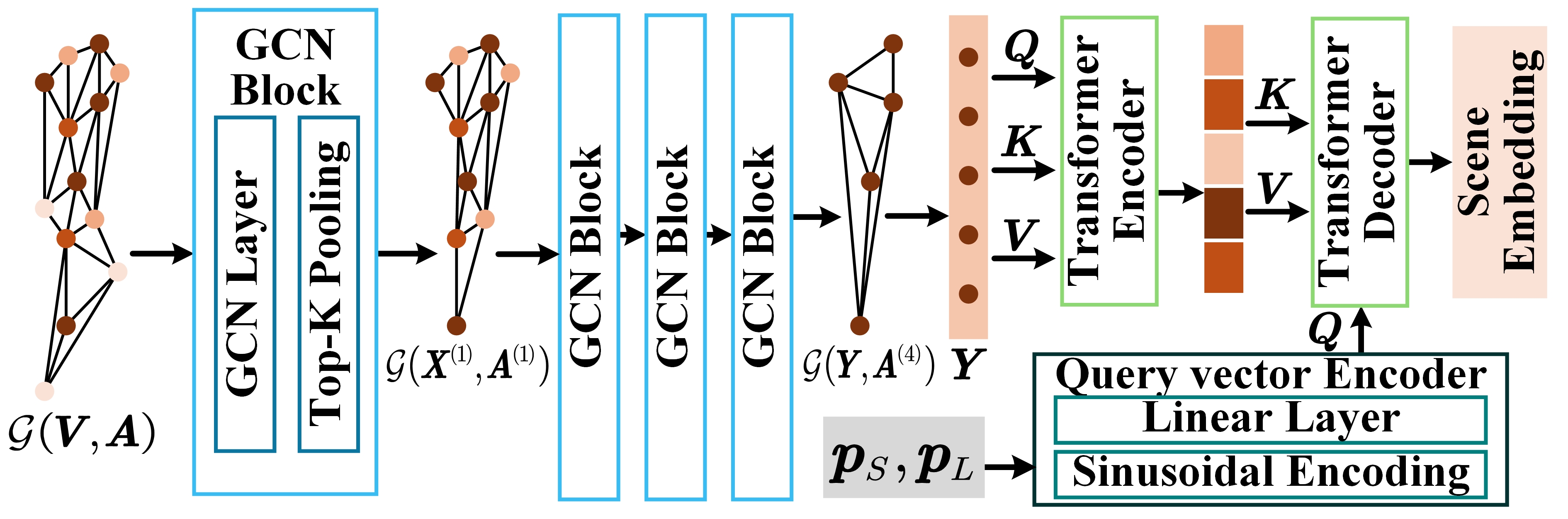}}
	\captionsetup{font=small, skip=2pt}
	\caption{\textbf{GCN-TF Encoder:} GCN blocks are used to simplify the graph and reduce the input sequence length for the Transformer. Then, the Transformer is applied to obtain the scene embedding.}
	\label{Encoder}
	\vspace{-4mm}
\end{figure}

\vspace{1mm}
\noindent \textbf{(1) GCN-TF Encoder:} The GCN-TF Encoder integrates a Graph Convolutional Network (GCN) and a Transformer (TF) to encode complex 3D scene meshes. The graph is first aggregated and simplified through $L$ GCN blocks. GCN layer is formulated as Eq.~\eqref{GCN},
\vspace{-2mm}
\begin{equation}
\boldsymbol{X}^{\left( l+1 \right)}=\sigma ( \hat{\boldsymbol{D}}^{-\small{\frac{1}{2}}}\hat{\boldsymbol{A}}\hat{\boldsymbol{D}}^{\small{-\frac{1}{2}}}\boldsymbol{X}^{\left( l \right)}\boldsymbol{W}^{\left( l \right)} ) \label{GCN} \vspace{-2mm}
\end{equation}
where diagonal matrix $\hat{\boldsymbol{D}}_{ii}=\sum_j{\hat{\boldsymbol{A}}_{ij}}$, $\boldsymbol{W}$ is a learnable matrix. We apply Top-K pooling\cite{gao2019graph} which selects the most informative vertices, after which adjacency relationships among the retained vertices are reconstructed. The process yields an encoded graph $\mathcal{G}_{Enc}(\boldsymbol{Y}, \boldsymbol{A}^{(4)})$ with substantially fewer vertices. The encoded vertex features $\boldsymbol{Y}$ form the input sequence of the Transformer encoder, where each vector $\boldsymbol{y}$ is treated as a token. In the decoder, the query matrix is derived from sinusoidally encoded source and listener coordinates, followed by a linear projection. The decoder queries the scene representation with these coordinates to extract position-specific scene information. The module structure is illustrated in Fig.~\ref{Encoder}.

\begin{figure}[tbp]
	\centerline{\includegraphics[scale=0.54]{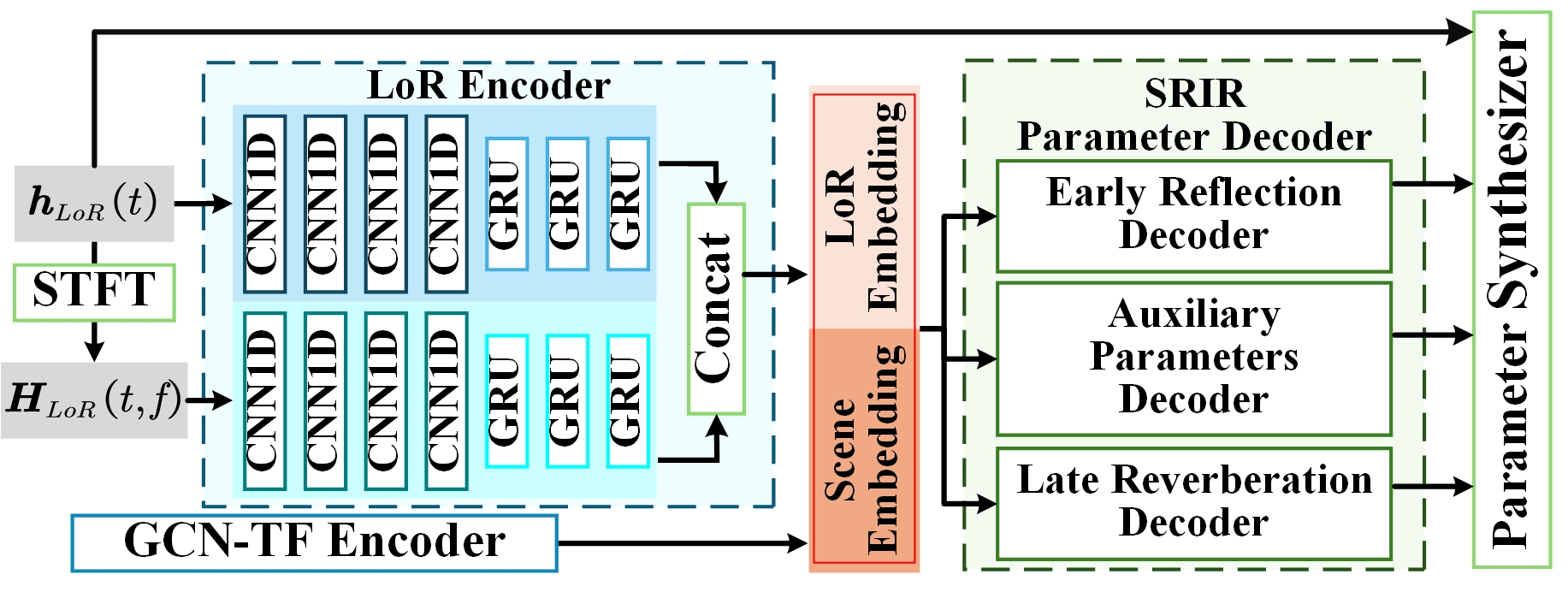}}
	\captionsetup{font=small, skip=2pt}
	\caption{The block diagram highlights the details of the LoR encoder and the SRIR parameter decoder. The output of the LoR encoder is concatenated with the output of the Scene Transformer and subsequently fed into the SRIR parameter decoder.}
	\label{ER-decoder}
	\vspace{-4mm}
\end{figure}

\vspace{1mm}
\noindent \textbf{(2) LoR Encoder:} This module embeds LoR information (see Fig.~\ref{ER-decoder}). To jointly capture temporal and spectral features, it adopts two parallel branches that take the waveform and Mel-spectrogram as inputs. Each branch consists of Convolutional Neural Networks (CNNs) for local feature extraction, followed by Gated Recurrent Units (GRUs) for temporal modeling.

\vspace{1mm}
\noindent \textbf{(3) SRIR Parameter Decoder and Parameter Synthesizer:} Following M2PAIR\cite{li2025m2pair}, the decoder predicts perceptual parameters of SRIR rather than the full SRIR. The decoder comprises three parallel modules: \textbf{1)} Early Reflection Decoder: outputs the energy-normalized waveform of early reflections (excluding LoR), $\boldsymbol{h}_{ER}^{\prime}$; \textbf{2)} Auxiliary Parameters Decoder: estimates SRIR duration $T_{60}$, early reflection energy $g_{ER}$, and late reverberation energy $g_{LR}$; \textbf{3)} Late Reverberation Decoder: generates the energy-normalized subband envelopes of late reverberation, $\boldsymbol{E}_{LR}$. The Parameter Synthesizer (PS) reconstructs each SRIR channel according to Eq.~\eqref{S1}\eqref{S2},
\vspace{-1mm}
\begin{equation}
	\boldsymbol{h}_S=g_{ER}\boldsymbol{h}_{ER}^{\prime}+\boldsymbol{h}_{LoR}+g_{LR}\boldsymbol{h}_{LR}^{\prime}
	\label{S1} \vspace{-1mm}
\end{equation}
\vspace{-3mm}
\begin{equation}
	\boldsymbol{h}_{LR}^{\prime}=\sum_F{\mathrm{Interp}.\left( \boldsymbol{e}_{LR,f},T_{60} \right) \cdot n_f}
	\label{S2} \vspace{-2mm}
\end{equation}
where, $\boldsymbol{h}_{LR}^{\prime}$ denotes the energy-normalized late reverberation, $\boldsymbol{e}_{LR,f}$ is the energy envelope of frequency band $f$, $\mathrm{Interp}.\left( \cdot ,t \right)$ is the interpolation operator that resamples the input to length $t$, and $n_f$ denotes band-limited noise in frequency band $f$.

\vspace{1mm}
\noindent \textbf{(4) Loss Function:}
Mean Absolute Error (MAE) loss is directly applied to the Auxiliary Parameters and Late Reverberation outputs. Due to the complexity of the early reflection waveform, its loss comprises the following components: \textbf{1)} Mel-spectrogram loss $\mathcal{L} _{\mathrm{Mel}}^{K,F}$, where $K$ and $F$ denote the total number of time frames and frequency bands, respectively; \textbf{2)} Waveform Mean Squared Error (MSE) loss $\mathcal{L} _{\mathrm{W}}$; \textbf{3)} Inter-channel waveform difference MSE loss $\mathcal{L} _{\mathrm{IC}}$, encouraging the model to capture relationships among SRIR channels. The complete loss function of early reflection waveform is shown in Eq.~\eqref{LossC},
\vspace{-1mm}
\begin{equation}
	\mathcal{L} =\alpha ( \mathcal{L} _{\mathrm{Mel}}^{K_1,F_1}+\mathcal{L} _{\mathrm{Mel}}^{K_2,F_2} ) +\beta \mathcal{L} _{\mathrm{W}}+\gamma \mathcal{L} _{\mathrm{IC}}
	\label{LossC} \vspace{-2mm}
\end{equation}
where $\alpha$, $\beta$, $\gamma$ are constants. Two different Mel-spectrogram resolutions are used to balance temporal and spectral resolution. The $\mathcal{L} _{\mathrm{IC}}$ formula is as Eq.~\eqref{LossIC},
\vspace{-2mm}
\begin{equation}
	\mathcal{L} _{\mathrm{IC}}=\frac{1}{CT}\sum_C{\left( \left( \boldsymbol{h}_{S}^{c}-\boldsymbol{h}_{S}^{c+1} \right) -\left( \hat{\boldsymbol{h}}_{S}^{c}-\hat{\boldsymbol{h}}_{S}^{c+1} \right) \right) ^2}
	\label{LossIC} \vspace{-3mm}
\end{equation}
where, $\hat{\cdot}$ represents the predicted value, $C$ denotes the number of channels and T denotes the number of samples.

\vspace{-1mm}
\subsection{Dataset}
\label{sec:Dataset}
\begin{figure}[t]
	\centering
	\begin{subfigure}[b]{0.48\linewidth}
		\centering
		\includegraphics[width=4cm]{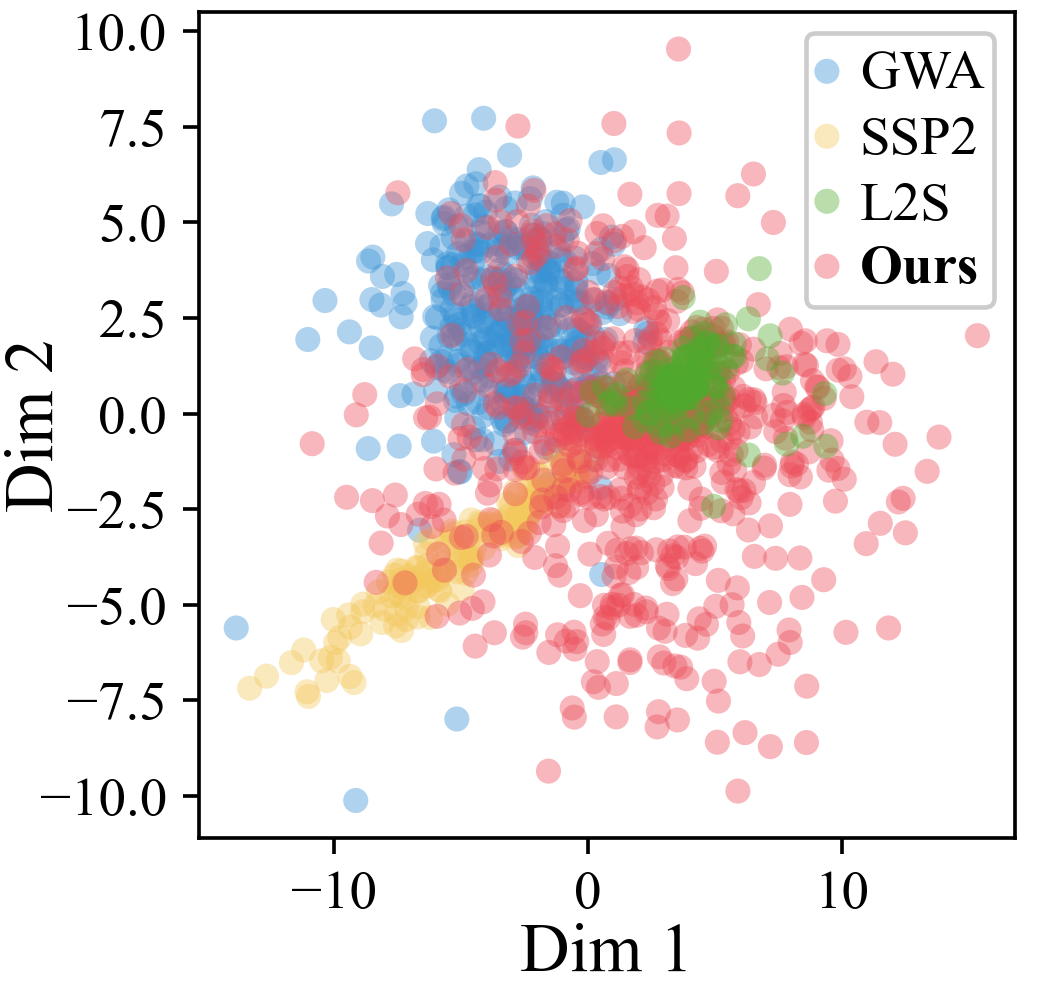}
		\vspace{-1mm}
		\caption{PCA-based dimensionality reduction of normalized energy spectra}
	\end{subfigure}
	\hfill
	\begin{subfigure}[b]{0.48\linewidth}
		\centering
		\includegraphics[width=4cm]{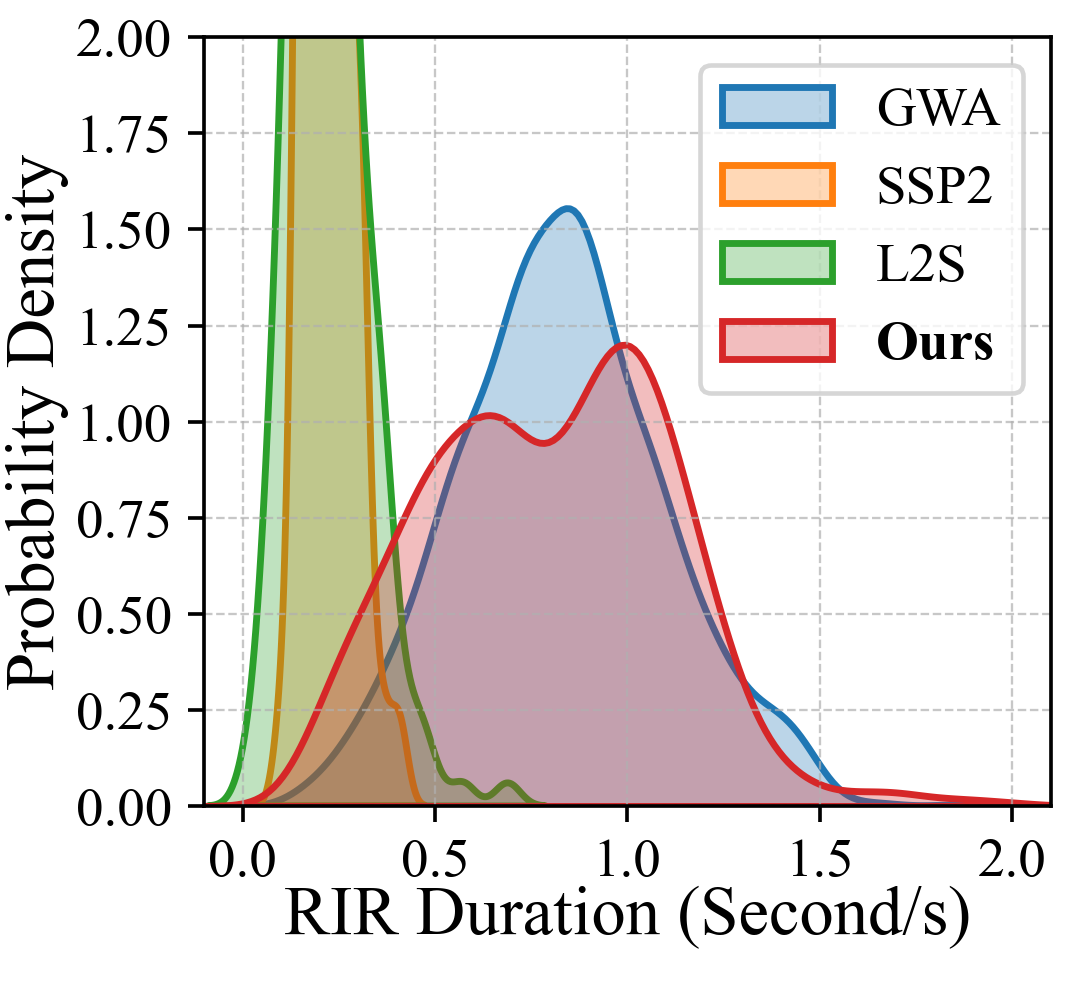}
		\vspace{-1mm}
		\caption{Histograms of SRIR duration distributions ($T_{60}$)}
	\end{subfigure}
	\vspace{-2mm}
	\caption{Feature distribution of RIRs across different datasets.}
	\label{Data}
	\vspace{-4mm}
\end{figure}

\vspace{-1mm}
To address the limitations of existing datasets, we constructed a new dataset based on GWA\cite{tang2022gwa} utilized in M2R. The dataset comprises $10^3$ residential scenes from 3D-FRONT\cite{fu20213d}, each with first-order Ambisonics A-format SRIRs simulated for $10^3$ source-listener coordinates using the pygsound (A GA toolbox)\cite{tang2020improving}. Mainstream datasets (GWA, L2S\cite{ratnarajah2024listen2scene}, Soundspaces 2 (SSP2)\cite{chen2022soundspaces}) mainly include residential environments with relatively homogeneous RIR characteristics, whereas VR applications demand more diverse acoustic conditions to provide richer perceptual cues. To address this, we varied reflection coefficients and scattering factors for reflective surfaces, significantly enhancing the perceptual diversity of RIRs and improving the distributional diversity of the dataset. As illustrated in Fig.~\ref{Data}, the resulting dataset demonstrates a substantially broader distribution in both energy spectra and durations.


\section{EXPERIMENT AND RESULTS}
\label{sec:EXPERIMENT}\vspace{-1mm}
\subsection{Benchmark Systems}
\label{sec:Benchmark}\vspace{-1mm}
\textbf{MESH2IR \cite{ratnarajah2022mesh2ir}:} This model takes scene geometry (without acoustic properties) together with source and listener coordinates as input, and outputs MRIRs. In this work, we modify its output channels to generate SRIRs. The model produces RIRs of length 4096, which, at a 48 kHz sampling rate, cover only early reflections.

\noindent \textbf{Listen2Scene \cite{ratnarajah2024listen2scene}:} The model architecture is essentially the same as M2R, but differs in input. L2S incorporates acoustic properties of the scene, specifically the reflectivity and scattering coefficients at 1 kHz. Experimental results indicate that using the 1 kHz band outperforms full-band input. In this study, the full-band variant is referred to as L2S-Full.

\noindent \textbf{M2PAIR \cite{li2025m2pair}:} With the same input as M2R, this model outputs MRIR perceptual parameters, which are then synthesized into high-quality MRIRs using signal processing. In this study, its output channels are adapted to generate SRIRs.

\vspace{-1mm}
\subsection{Inference Result Evaluation}
\label{sec:obj_err}\vspace{-1mm}
To ensure fair comparison, SRIRs were truncated for models limited to 4096 samples (M2R and L2S), denoted as “$^-$”, while our model and M2PAIR output complete SRIRs. Objective evaluation metrics include SRIR waveform MAE ($10^{-4}$), reverberation time ($T_{60}$/s), total energy (En./dB), direct-to-reverberant ratio (DRR/dB), and Mel-spectrogram errors (Mel/dB for high-frequency resolution and Mel-T/dB for high temporal resolution). Tab.~\ref{RIR-err} reports the errors of different models, showing that our model consistently achieves superior performance.

During SRIR synthesis, our model directly adds the input LoR to the remaining components (Eq.~\eqref{S1}), such that the LoR in the output is groundtruth. To further assess accuracy, we trained other models on SRIRs with LoR removed and evaluated them on LoR-free SRIRs (Tab.~\ref{reverb-err}). DRR was omitted in these cases due to its strong dependence on LoR. Results indicate that even on LoR-free SRIRs, our model maintains superior performance. The key architectural innovation is the introduction of LoR as auxiliary modality and the corresponding LoR encoder. Ablation studies show that removing this module (w/o LoR) consistently degrades performance across all metrics.

\begin{table}[tbp]
	\begin{center}
		\small
		\begin{tabular}{p{1.5cm}|C{0.6cm}C{0.6cm}C{0.6cm}C{0.6cm}C{0.6cm}C{1cm}}
			\hline
			\multicolumn{1}{l|}{}  & \textbf{MAE}  & \textbf{$\boldsymbol{T}_{\mathbf{6}\mathbf{0}}$}  & \textbf{En.} & \textbf{DRR}  & \textbf{Mel}  & \textbf{Mel-T} \\ \hline
			M2PAIR                 & 0.81          & 0.28          & 7.55            & 11.38         & 8.59          & 4.35           \\
			w/o LoR       & 0.69          & 0.28          & 6.26            & 6.30          & 8.46          & 4.25           \\
			\textbf{Ours}          & \textbf{0.55} & \textbf{0.26} & \textbf{3.98}   & \textbf{5.09} & \textbf{7.27} & \textbf{3.52}  \\ \hline
			M2R                    & 4.01          & -             & 6.92            & 9.85          & 12.89         & 9.60           \\
			L2S               & 8.66          & -             & 10.28           & 9.04          & 11.78         & 9.95           \\
			L2S-Full               & 3.52         & -             & 6.45            & 9.32          & 10.77         & 8.43          \\
			M2PAIR$^-$                 & 5.00          & -             & 7.79            & 10.79         & 10.73         & 9.05           \\
			w/o LoR$^-$      & 3.61          & -             & 6.19            & 5.83          & 9.28          & 7.68           \\
			\textbf{Ours$^-$}    & \textbf{2.05} & \textbf{-}    & \textbf{3.60}   & \textbf{4.74} & \textbf{5.61} & \textbf{4.40}  \\ \hline
		\end{tabular}
	\end{center}
	\vspace{-6mm}
	\caption{\textbf{Full SRIR computation error}, where the upper part corresponds to complete SRIRs and the lower part to truncated SRIRs.}
	\label{RIR-err}
	\vspace{-1mm}
\end{table}

\begin{table}[tbp]
	\vspace{-1mm}
	\begin{center}
		\small
		\begin{tabular}{p{1.5cm}|C{0.8cm}C{0.8cm}C{0.8cm}C{0.8cm}C{1cm}}
			\hline
			\multicolumn{1}{l|}{} & \textbf{MAE}  & \textbf{$\boldsymbol{T}_{\mathbf{6}\mathbf{0}}$}  & \textbf{En.}  & \textbf{Mel}  & \textbf{Mel-T} \\ \hline
			M2PAIR                & 0.76          & 0.29          & 8.56          & 9.45          & 4.83           \\
			w/o LoR               & 0.69          & 0.28          & 7.71          & 8.54          & 4.31           \\
			\textbf{Ours}         & \textbf{0.55} & \textbf{0.26} & \textbf{4.41} & \textbf{7.34} & \textbf{3.57}  \\ \hline
			M2R                   & 3.68          & -             & 21.15         & 18.69         & 16.85          \\
			L2S                   & 2.80          & -             & 11.96         & 14.30         & 10.79          \\
			L2S-Full              & 3.17          & -             & 11.16         & 13.71         & 10.21          \\
			M2PAIR$^-$                & 3.87          & -             & 8.69          & 11.07         & 9.26           \\
			w/o LoR$^-$              & 3.61          & -             & 7.88          & 10.17         & 8.32           \\
			\textbf{Ours$^-$}        & \textbf{2.05} & \textbf{-}    & \textbf{3.99} & \textbf{6.35} & \textbf{4.94}  \\ \hline
		\end{tabular}
	\end{center}
	\vspace{-6mm}
	\caption{\textbf{LoR-free SRIR computation error.}}
	\vspace{-5mm}
	\label{reverb-err}
\end{table}

%
\noindent \textbf{Dataset Diversity Verification:} The L2S study reported that using acoustic properties at 1 kHz as input yielded optimal performance \cite{ratnarajah2024listen2scene}. This result partially contradicts physical acoustics principles \cite{kuttruff2016room} and may result from the limited diversity of the L2S dataset, whose RIR feature distribution is relatively concentrated. Tab.~\ref{RIR-err} and Tab.~\ref{reverb-err} indicate that L2S-Full outperforms L2S, which in turn surpasses M2R, further confirming the superior diversity of our dataset.

\vspace{-2mm}
\subsection{Computational Complexity}
\label{sec:speed}\vspace{-1mm}
Computational complexity is evaluated in terms of model parameters (Params., $10^6$), floating-point operations per second (FLOPs, $10^6$), and actual computation time ($10^{-3}$ s). In practical applications, scene encoding triggered by scene transitions occurs far less frequently than SRIR decoding due to source-listener coordinate changes. Accordingly, both our method and benchmark systems are divided into static and dynamic components, with the dynamic component being critical for real-time performance. As shown in Tab.~\ref{speed}, the proposed model demonstrates far superior time efficiency to traditional methods. Compared with DL-based benchmark systems, its complexity is relatively higher but still fully meets the real-time requirements of the target application. The computational efficiency of M2R, L2S, and L2S-Full is nearly identical; therefore, only M2R is reported in the table.

\vspace{-1mm}
\subsection{Subjective Evaluation}
\label{sec:subj}\vspace{-1mm}
We conducted a subjective evaluation following MUSHRA \cite{ITU-R_BS.1534-3}, using groundtruth SRIRs as the reference. Participants rated the perceptual similarity between the test and reference audio on a 10-point scale, with 10 indicating complete similarity. As SRIRs represent system functions, they were convolved with the audio signals. Fifteen test samples covering speech, music, and songs were evaluated by ten participants. Results, presented in Tab.~\ref{subj}, indicate that our model achieved the highest perceptual scores.

\begin{table}[tbp]
	\begin{center}
		\small
		\begin{tabular}{ll|lll}
			\hline
			& \textbf{}       & \textbf{Params.} & \textbf{FLOPs} & \textbf{Times} \\ \hline
			(i)   & GA              & -                & -              & 6942.15        \\ \hline
			(ii)  & M2R-Static      & 0.0012           & 0.0012         & 14.96          \\
			& \textbf{Ours-Static}     & 25.19            & 6450.93        & 18.33          \\ \hline
			(iii) & M2R-Dynamic     & 115.31           & 23485.93       & 5.17           \\
			& M2PAIR-Dynamic  & 27.80            & 208.20         & 89.02          \\
			& w/o LoR-Dynamic & 292.80           & 9308.63        & 450.22         \\
			& \textbf{Ours-Dynamic}    & 329.76           & 10742.56       & 485.49         \\ \hline
			(iv)  & GA-LoR          & -                & -              & 310.09         \\
			& DL-model        & 329.76           & 10742.56       & 88.97          \\
			& PS              & -                & -              & 86.43          \\ \hline
		\end{tabular}
	\end{center}
	\vspace{-6mm}
	\caption{\textbf{Table of Computational Complexity}, consisting of four sections: (i) traditional methods, (ii) static components of deep learning models, (iii) dynamic components, and (iv) the individual modules of this study, where, GA-LoR refers to the computation of LoR based on the GA method.}
	\vspace{-2mm}
	\label{speed}
\end{table}

\begin{table}[tbp]
	\begin{center}
		\small
		\begin{tabular}{l|llll}
			\hline
			& PS   & M2PAIR & w/o LoR & \textbf{Ours} \\ \hline
			\textbf{Mean} $\uparrow$ & 8.08 & 5.12   & 5.71   & \textbf{7.04} \\
			\textbf{Var} $\downarrow$  & 3.27 & 5.22   & 5.18   & \textbf{3.13} \\ \hline
		\end{tabular}
	\end{center}
	\vspace{-6mm}
	\caption{\textbf{Subjective test scores (mean and variance),} where PS refers to the result obtained by directly feeding the perceptual parameters of the ground truth into the PS module for decoding. The purpose is to illustrate the impact of the parameterized SRIR encoding-decoding process on perceptual quality.}
	\label{subj}
	\vspace{-3mm}
\end{table}

\vspace{-1mm}
\section{Conclusion and Future Work}
\label{sec:conclusion}\vspace{-2mm}
This study addresses the challenge of auralization in VR scenarios. We propose a scene-waveform multimodal model that computes SRIRs in real time from scene geometry, acoustic properties, source-listener coordinates, and LoR waveform. For the first time, LoR is incorporated as auxiliary modality to enhance model performance, and a novel SRIR dataset is constructed. The dataset exhibits greater diversity and provides SRIRs with richer characteristics. Experimental results demonstrate superior performance, with LoR markedly enhancing SRIR quality, and the model’s computational speed meets the real-time requirements of VR auralization. In future work, we plan to explore alignment strategies between scene and waveform modalities and to conduct more extensive subjective evaluations.

\vfill\pagebreak


\bibliographystyle{IEEEbib}
\bibliography{refs}

\end{document}